\documentclass[preprint,showpacs,showkeys,preprintnumbers,amsmath,amssymb,aps]{revtex4}
\usepackage{epsfig}
\usepackage{graphicx}% Include figure files
\usepackage{dcolumn}% Align table columns on decimal point
\usepackage{bm}% bold math
\usepackage{tabularx}

\def\bsub{\begin{subequations}}

\def\esub{\end{subequations}}

\begin{document}
\baselineskip 0.7cm

\title{   Constraints on Astro-unparticle Physics from SN 1987A} 

\author{Sukanta Dutta$^1$}\email{Sukanta.Dutta@fnal.gov}

\author{Ashok Goyal$^2$} \email{agoyal@iucaa.ernet.in}

\affiliation{$^1$ Department of Physics and Electronics, SGTB Khalsa College, University of Delhi. Delhi-110007. India.}

\author{  }

\affiliation{$^2$Department of Physics and Astrophysics, University of Delhi. Delhi-110007. India.}

\begin{abstract}

SN 1987A observations have been used to place constraints on the interactions between  standard model particles and unparticles.  In this study we calculate the energy loss from the supernovae core through scalar, pseudo scalar, vector, pseudo vector unparticle emission from nuclear bremsstrahlung for degenerate nuclear matter interacting through one pion exchange.  In order to examine the constraints on $d_{\cal U}=1$ we considered the emission of  scalar, pseudo scalar, vector, pseudo vector  and tensor  through the pair annihilation process $e^+e^-\to {\cal U}\,\gamma $.  In addition we have re-examined other pair annihilation processes.  The  most stringent bounds on the  dimensionless coupling constants for $d_{\cal U} =1$ and $\Lambda_{\cal U}= m_Z$ are obtained from nuclear bremsstrahlung process for the pseudo scalar and pseudo-vector couplings   $\bigl\vert\lambda^{\cal P}_{0,1}\bigr\vert\leq 4\times 10^{-11}$ and for  tensor interaction, the best limit on dimensionless coupling is obtained from $e^+\, e^-\rightarrow {\cal U}\,\gamma$ and we get $\bigl\vert\lambda^{\cal T}\bigr\vert \leq 6\times 10^{-6}$. 

\end{abstract}



\maketitle

\section{Introduction}

\label{introduction}
Recently Georgi \cite{georgi, georgi1} has considered the interesting possibility of the existence of the new physics above the TeV scale through the introduction of Unparticles. In this scheme at high energies   there is hidden sector with a non trivial IR fixed point $\Lambda_{\cal U} $, below which there is scale invariance. At energies above $\Lambda_{\cal U} $, there is a hidden sector operator ${\cal O}_{\cal UV}$ of dimension $d_{\cal UV}$ that couples to SM operator ${\cal O}_{SM}$ of dimension $n$ through the exchange of high mass $M_{\cal U}$ particles
\begin{eqnarray}
{\cal L}_{\cal UV}&=&\frac{{\cal O}_{SM}\,\, {\cal O}_{\cal UV}}{M^{d_{\cal UV} +n-4}}
\end{eqnarray}
Below $\Lambda_{\cal U} $, the hidden sector  becomes scale invariant and the operator ${\cal O}_{UV}$ goes over to ${\cal O}_{\cal U} $ an unparticle operator of dimension $d_{\cal U}$.
\begin{eqnarray}
{\cal L}_{\cal U}&=&{\cal C_U}\,\,\frac{\Lambda_{\cal U}^{d_{\cal UV}-d_{\cal U}}}{M^{d_{\cal UV} +n-4}\,\,}{\cal O}_{SM}\,\, {\cal O}_{\cal UV}
\end{eqnarray}
where ${\cal C_U} $ is the dimensionless coupling constant and the phase space of ${\cal O_U}$ is the same as the phase space of  a massless  particles and $d_{\cal U}$ is free to take non integer values. The unparticle phase space for $d_{\cal U}$ dimensions is then given by
\begin{equation}
{\cal A}_{d_{\cal U}u}= \frac{16\,\pi^{5/2}}{\left(2\,\pi\right)^{2\,d_{\cal U}}}\,\,\,\frac{\Gamma \left(d_{\cal U}+ \frac{1}{ 2}\right)}
{\Gamma \left(d_{\cal U} -1 \right)\,\,\,\Gamma \left(2\,\,d_{\cal U}\right)}
\end{equation}
which reduces to the standard value for $d_{\cal U}$=1. Since the unparticle sector appears at low energy as massless fields coupled to SM particles very weakly, their emission from the stellar matter will result in energy loss and thus to the cooling of these objects. This can be used for putting constraints on the parameters of the theory.  Recently Davoudiasl \cite{davoudiasl} , Hannestad {\it el. al. } \cite{hannestad} and Freitas \& Wyler \cite{freitas} have used astrophysical limits to constrain the unparticle physics. Davoudiasl \cite{davoudiasl} , Hannestad {\it el. al. } \cite{hannestad} essentially used dimensional analysis considerations to put the constraints on vector unparticles from supernova SN 1987 A.  Freitas \& Wyler \cite{freitas} extended the analysis to include scalar and pseudo-scalar and axial unparticle operators. They however, did not consider the scalar and pseudo-scalar couplings like $\bar f\,f\, {\cal O_U^S}$ and $\bar f\,\gamma_5\,f\, {\cal O_U^P}$.
These authors considered  the dominant nucleon bremsstrahlung namely $NN\rightarrow N\,N\,{\cal U}$ process for the emission of unparticles from the supernova core.  They  have estimated the energy loss due to nucleon bremsstrahlung by factorizing the process into a \lq hard\rq \,\, $NN$ collision process and \lq soft\rq \,\, unparticle emission from one of the external nucleon. They then calculated the rates in non- relativistic limit.  
\par In addition the authors of reference \cite{hannestad} also considered the vector unparticle production through electron neutrino - anti neutrino annihilation. Recently Lewis  \cite{lewis} extended these calculations to include energy loss rates for tensor particle production from photon - photon ($\gamma \gamma \rightarrow {\cal U}$) and electron - positron ($e^+\, e^- \rightarrow {\cal U}$) annihilation. It is obvious that pair annihilation will not give any constraints for $d_{\cal } =1$ simply because energy momentum conservation forbids pair annihilation to a single massless particle. The most restrictive bounds from these studies have been obtained for vector unparticles \cite{hannestad} and the constraints on  scalar unparticles \cite{freitas} are much weaker. 
\par The unparticles can arise as stated in \cite{georgi}  from the hidden sector or from strongly interacting magnetic phase of a specific class of supersymmetric theories \cite{Fox:2007sy} or from hidden valleys model \cite{strassler}. However, we also note that under a specific conformal invariance \cite{intriligator} the  propagators for vector and tensor are modified. Fox {\it et. al. } \cite{Fox:2007sy} from a study of supersymmetric QCD in the conformal regime have shown that  the interaction of dimension $d_{\cal U}<2$ unparticles with SM Higgs break conformal invariance once  the Higgs acquire non - zero VEV. The theory becomes non-conformal  below the scale and unparticle physics loses its relevance.  In this study we assume that the conformal invariance continues to remain valid down to the energy relevant in supernovae processes and that astrophysical constraints would keep in the unfolding issues in unparticle physics. This  is also in conformity with the view taken in references \cite{davoudiasl,hannestad,freitas,lewis}. 
\par Recently there has been a lot of interest in  phenomenological studies of
 unparticles \cite{Huitu:2007im,Cakir:2007dz,Alan:2007ui,Cheung:2007ue,
 Luo:2007bq, Chen:2007vv,Ding:2007bm,Liao:2007bx,Aliev:2007qw,Li:2007by,
Duraisamy:2007aw,Lu:2007mx,Greiner:2007hr,Choudhury:2007js, Chen:2007qr,
Aliev:2007gr,Mathews:2007hr,Zhou:2007zq,Ding:2007zw,Chen:2007je,Bander:2007nd,
Rizzo:2007xr,Cheung:2007ap,Chen:2007zy,Zwicky:2007vv,Kikuchi:2007qd,
Mohanta:2007ad,Huang:2007ax,Lenz:2007nj,Choudhury:2007cq,Zhang:2007ih,Li:2007kj,
Deshpande:2007jy,Mohanta:2007uu,Cacciapaglia:2007jq,Neubert:2007kh,Luo:2007me,
Bhattacharyya:2007pi,Majumdar:2007mp,Alan:2007ss,Chen:2007pu,Hur:2007cr,
Anchordoqui:2007dp,Balantekin:2007eg,Aliev:2007rm,Iltan:2007ve,Chen:2007cz,
Alan:2007rg,Delgado:2007dx,Sahin:2007pj,Majhi:2007tu,Kumar:2007af,Ding:2007jr,
Kobakhidze:2007zs}. The  astrophysical studies were performed in references 
\cite{Das:2007nu, Liao:2007ic,Deshpande:2007mf,Das:2007cc} where 
it is assumed that conformal invariance holds down to low energy regime relevant  for the processes  in the supernova. Studies were also carried out on the impact of unparticles in the cosmology in references \cite{davoudiasl,Kikuchi:2007az,Alberghi:2007vc,McDonald:2007bt}.
In this paper we revisit the energy loss rate from the emission of unparticles from the supernova core for the dominant nucleon bremsstrahlung and subdominant pair annihilation processes. For the purpose we take vector, axial vector, scalar and pseudo-scalar unparticles. Energy loss rate due to emission of unparticles from Supernova core through the nucleon bremsstrahlung process is calculated by taking a more traditional route which has well served in estimating the energy loss rate  from supernova core for the case of neutrino and  axion emission \cite{friman,Iwamoto:1984ir, turner, Iwamoto:1989mh}. In these studies the matrix element has been calculated by assuming $N\, N$ interaction through single pion exchange through the standard coupling 
\begin{equation}
{\cal L} = -\, \frac {i\, f}{m_\pi}\,\, \bar N \,\gamma^\mu\,\gamma_5\,\vec \tau\, N\,\,\cdot\,\, (\partial_\mu\, \vec \pi )
\end{equation}
(Here $f$ is the dimensionless coupling constant of order 1) and through the radiation of weakly interacting light particle from any of the four nucleon legs. 
\par We also calculate energy loss rate due to unparticle production through pair annihilation processes for couplings mentioned above. In order to obtain a bound for $d_{\cal U}=1$ case, we consider the pair annihilation of charged leptons through $e^+e^-\rightarrow {\cal U}\, \gamma$. This process was mentioned by author of reference  \cite{hannestad} as a possible competitive process for obtaining constraints for $d_{\cal U}=1$. In section \ref{nucleonbrem},  we list the effective interaction between scalar , pseudo-scalar , vector , axial vector and tensor unparticles with the SM fields and calculate the energy loss rate from nucleon bremsstrahlung process. In section \ref{eeannihil} we calculate the pair annihilation to unparticles and the resulting energy loss. In section \ref{discuss} we numerically evaluate the energy loss from the supernova core and put constraints from SN 1987A on unparticle couplings with SM particles and discuss the results.
\section{Unparticle Emission from Nucleon Bremsstrahlung Process}
\label{nucleonbrem}
The effective scalar and pseudo-scalar unparticle interaction with SM particles under the present study are 
\begin{eqnarray}
\frac{\lambda_0^{\cal S}}{\Lambda_{\cal U}^{d_{\cal U}-1}}\,\, \bar f\,\,  f\,\, {\cal O_U}\, ; \, \,\, \frac{\lambda_0^{\cal P}}{\Lambda_{\cal U}^{d_{\cal U}-1}}\,\, \bar f\,\gamma_5\,  f\,\, {\cal O_U}\, ;\,\,\, \frac{\lambda_0^{{\cal S}_1}}{\Lambda_{\cal U}^{d_{\cal U}}}\,\, \bar f\,\gamma^\mu \, f\,\,\, \bigl(\partial_\mu\, {\cal O_U}\bigr)\, ;\frac{\lambda_0^{{\cal P}_1}}{\Lambda_{\cal U}^{d_{\cal U}}}\,\, \bar f\,\gamma_\mu\,\gamma_5 \, f\, \bigl(\partial_\mu\, {\cal O_U}\bigr) \, {\rm and}\,\, \frac{\lambda_0^\gamma}{\Lambda_{\cal U}^{d_{\cal U}}}\,\, {\cal F}^{\mu\nu}{\cal F}_{\mu\nu}\, {\cal O_U}\nonumber\\ \label{scalcoup}
\end{eqnarray}
For the vector and axial vector the unparticle operators, we have
\begin{eqnarray}
 \frac{\lambda_1^{\cal V}}{\Lambda_{\cal U}^{d_{\cal U}-1}}\,\, \bar f\,\gamma_\mu \, f\,\,\,  {\cal O_U}^\mu\, ;\,\,\,{\rm and}\,\,\,  
 \frac{\lambda_1^{\cal A}}{\Lambda_{\cal U}^{d_{\cal U}-1}}\,\, \bar f\,\gamma_\mu\,\gamma_5 \, f\, {\cal O_U}^\mu\label{veccoup}
\end{eqnarray}
and for tensor unparticles the interactions are
\begin{eqnarray}
\frac{-\,i}{4}\frac{\lambda_{\cal T}} {\Lambda_{\cal U}^{d_{\cal U}}} \bar f \,\, \bigl( \gamma_\mu \stackrel {\leftrightarrow } {D}_\nu + \gamma_\nu \stackrel {\leftrightarrow}{D}_\mu \bigr) \,\,\psi_f\,\, {\cal O}_{\cal U}^{\mu\nu} \,\, {\rm and}\,\, 
\frac{\lambda_{\cal T}} {\Lambda_{\cal U}^{d_{\cal U}}} 
{\cal F}_{\mu\alpha}{\cal F}_{\nu}^\alpha\, {\cal O_U}^{\mu\nu}\label{tenscoup}
\end{eqnarray}
Here the dimensionless coupling constants $\lambda_i$ are related to the coupling constant ${\cal C_U}$ and the mass scale $M_{\cal U}$ through 
\begin{eqnarray}
\frac{\lambda_1^{\cal V,\, A}}{\Lambda_{\cal U}^{d_{\cal U}-1}}=
\frac{\lambda_0^{\cal S,\, P}}{\Lambda_{\cal U}^{d_{\cal U}-1}}={\cal C_U^{S,\,P,\,V,\,A}}\,\, \frac{\Lambda_{\cal U}^{3-d_{\cal U}}}{M^2_{\cal U}}\,\,&{\rm and}&\,\, 
\frac{\lambda_0^{{\cal S}_1,\, {\cal P}_1,\, {\cal T}}}{\Lambda_{\cal U}^{d_{\cal U}}}={\cal C_U}^{{\cal S}_1,\,{\cal P}_1,\, {\cal T}}\,\, \frac{\Lambda_{\cal U}^{2-d_{\cal U}}}{M^2_{\cal U}}
\end{eqnarray}
\par We now calculate the energy loss rate due to neutron bremsstrahlung 
\begin{equation}
N(p_1) + N(p_2) \rightarrow N(p_3) + N(p_4) + {\cal U}(P)
\end{equation}.
for the interaction considered above. The energy loss rate is given by
\begin{eqnarray}
\stackrel{\bf .}{\epsilon}_{\cal U}&=& {\cal A}_{d_{\cal U}} \int\left[\prod _{i=1}^4 \,\frac{d^3p_i}{2 E_i (2\pi)^3}\right] \Theta \bigl(P_0\bigr) \Theta\bigl( P^2\bigr) \bigl( P^2\bigr)^{d_{\cal U}-2} P_0\frac{1}{4}\, \sum \bigl\vert {\cal M}\bigr\vert^2\,f_1\,f_2\, (1-f_3)\, (1-f_4)
\end{eqnarray}
where $\sum \bigl\vert {\cal M}\bigr\vert^2 $ is the matrix element squared summed over spins and (1/4) is the statistical factor for identical neutrons and $f_i$'s are the Fermi -Dirac distribution functions. Introducing 
\begin{eqnarray}
1 = \int d^4P\,\, \delta^4\bigl( p_1+p_2-p_3-p_4-P\bigr)
\end{eqnarray}
and integrating over $d\bigl\vert \vec P\bigr\vert$, we get
\begin{eqnarray}
\stackrel{\bf .}{\epsilon}_{\cal U}&=& \frac{\pi^{5/2}}{(2\pi)^{2d_{\cal U}}}\,\frac{\Gamma (1/2)}{\Gamma (2\, d_{\cal U})} \left[\int\prod _{i=1}^4 \, \frac{d^3\vec p_i}{2 E_i (2\pi)^3}\right]  \bigl(P_0\bigr)^{2\,d_{\cal U}}\,dP_0\, d\Omega_P\,\frac{1}{4} \sum \bigl\vert {\cal M}\bigr\vert^2\,f_1\,f_2\, (1-f_3) (1-f_4)\nonumber\\
\end{eqnarray}
\par Since the supernovae temperature $\approx $ 30 MeV  is small compared to the nucleon mass, the non-relativistic treatment  of nucleons is adequate. In the limit of treating the nucleon propagator non - relativistically and keeping only the leading term namely
$$\frac {i}{(p+q)^2-m_N^2}\rightarrow \frac {i}{\pm \,2\, m_N\omega}$$
where $\omega $ is the energy of the emitted unparticle. The leading  contribution  from the  diagrams when the scalar and vector unparticles are emitted from the outgoing and incoming nucleon legs respectively, cancels  in pairs. This has also been emphasized by the author of the reference \cite{hannestad}, who then use next to the leading order term namely, the quadrupole contribution for the vector case.
\par The matrix element squared for the scalar, pseudoscalar (axion) and neutrino emission has been calculated in the references \cite{friman,Iwamoto:1984ir, turner, Iwamoto:1989mh,Ishizuka:1989ts}. For the vector and the axial vector case, the squared matrix element  can be calculated by a slight modification of their result. The angular integrations can be done by neglecting pion mass in comparison with the nucleon Fermi momenta which is typically of the order of 356 MeV. This results in not more than 10-15\% change in the energy loss. By using the standard techniques of replacing the neutron momenta by their Fermi momenta wherever possible, we get 
\begin{figure}[htb]
\begin{center}
\vskip -5 cm
\includegraphics[width=15cm,height=30cm]{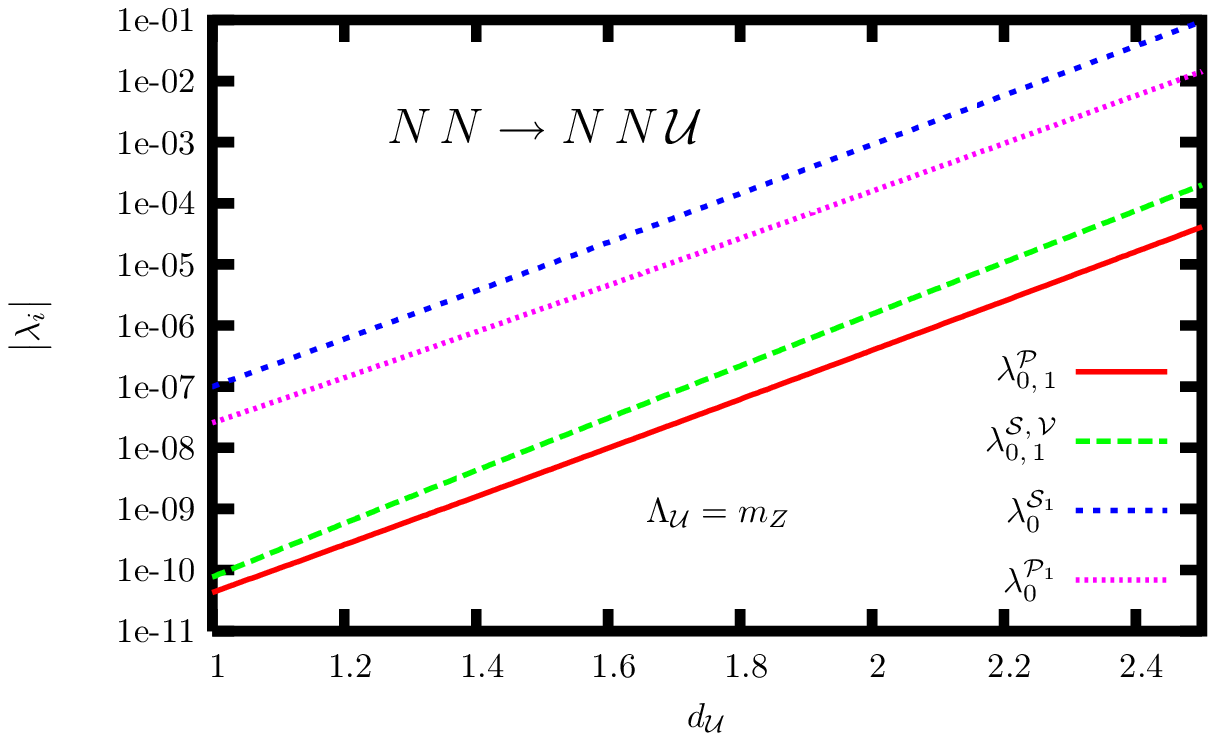}
\vskip -18 cm 
\caption{Contour on $d_{\cal U}$ and coupling $\bigl\vert \lambda_i\bigr\vert$ plane for $\Lambda_{\cal U}= m_Z$ for the unparticle emitting nucleon bremsstrahlung processes}
\label{bremfig}
\end{center}
\end{figure}
\begin{eqnarray}
\stackrel{\bf .}{\epsilon}_{\cal U}\bigl({\cal P}\bigr)&\approx& \frac{64}{(2\pi)^{2d_{\cal U}+5}}\, \frac{1} {\Gamma (2\, d_{\cal U})}\,\, T^{(2\, d_{\cal U} +4)}\, {m_n^\star}^2\,\, p_F\,\left(\frac{\bigl\vert\lambda^{\cal P}_{0,\,1}\bigr\vert}{\Lambda_{\cal U}^{d_{\cal U}-1}}\right)^2\left(\frac{f}{m_\pi}\right)^4\,\, {\cal J}(1)\\
&&\,\, {\rm where}\,\,\, {\cal J}(n) = \pi^2\int_{0}^{\infty}\frac{ y^{2\,d_{\cal U}+n}}{e^y-1}\,\, \left(\frac{2}{3}+\frac{y^2}{6\,\pi^2}\right)\,\, dy
\end{eqnarray}
For $d_{\cal U}$ = 1, it reduces to the well known axion case namely
\begin{eqnarray}
\stackrel{\bf .}{\epsilon}_{\cal U}({\cal P})\bigg\vert_{\bigl(d_{\cal U}=1\bigr)}&\approx& \frac{31}{1890\,\pi }\,\,\bigl\vert\lambda^P_{0,\,1}\bigr\vert^2\,\, \left(\frac{f}{m_\pi}\right)^4\,{m_n^\star}^2\,\, p_F\,\, T^6
\end{eqnarray}
For the scalar and vector case we get  
\begin{eqnarray}
\stackrel{\bf .}{\epsilon}_{\cal U}\bigl({\cal S,\,V}\bigr)&\approx & \frac{256}{75\,(2\pi)^{(2\,d_{\cal U}+5)} }\, \frac{1} {\Gamma (2\, d_{\cal U})}\,\left(\frac{\bigl\vert\lambda^{S,\,V}_{0,\,1}\bigr\vert}{\Lambda_{\cal U}^{d_{\cal U}-1}}\right)^2\left(\frac{f}{m_\pi}\right)^4\, p_F^5 \,\, T^{(2\,d_{\cal U}+2)}\,\, {\cal J}(-1)
\end{eqnarray}
For $d_{\cal U}$ = 1, we recover the Ishizuka and Yoshimura's \cite{Ishizuka:1989ts} result for the scalar dilaton emission. For the other two cases, we get 
\begin{eqnarray}
\stackrel{\bf .}{\epsilon}_{\cal U}\bigl({\cal P}_1\bigr)&\approx & \frac{64}{(2\pi)^{(2\,d_{\cal U}+5)} }\, \frac{1} {\Gamma (2\, d_{\cal U})}\,{m_n^\star}^2\left(\frac{\bigl\vert\lambda^{{\cal P}_1}_{0}\bigr\vert}{\Lambda_{\cal U}^{d_{\cal U}}}\right)^2\left(\frac{f}{m_\pi}\right)^4\, p_F \,\, T^{(2\,d_{\cal U}+6)}\,\, {\cal J}(3) \,\, \, {\rm and}\\
\stackrel{\bf .}{\epsilon}_{\cal U}\bigl({\cal S}_1\bigr)&\approx & \frac{256}{75\,(2\pi)^{(2\,d_{\cal U}+5)} }\, \frac{1} {\Gamma (2\, d_{\cal U})}\left(\frac{\bigl\vert\lambda^{{\cal S}_1}_{0}\bigr\vert}{\Lambda_{\cal U}^{d_{\cal U}}}\right)^2\left(\frac{f}{m_\pi}\right)^4\, p_F^5 \,\, T^{(2\,d_{\cal U}+4)}\,\, {\cal J}(1) 
\end{eqnarray}
\section{Unparticle Emission from Pair Annihilation}
\label{eeannihil}
\begin{figure}[htb]
\begin{center}
\vskip -5 cm
\includegraphics[width=15cm,height=30cm]{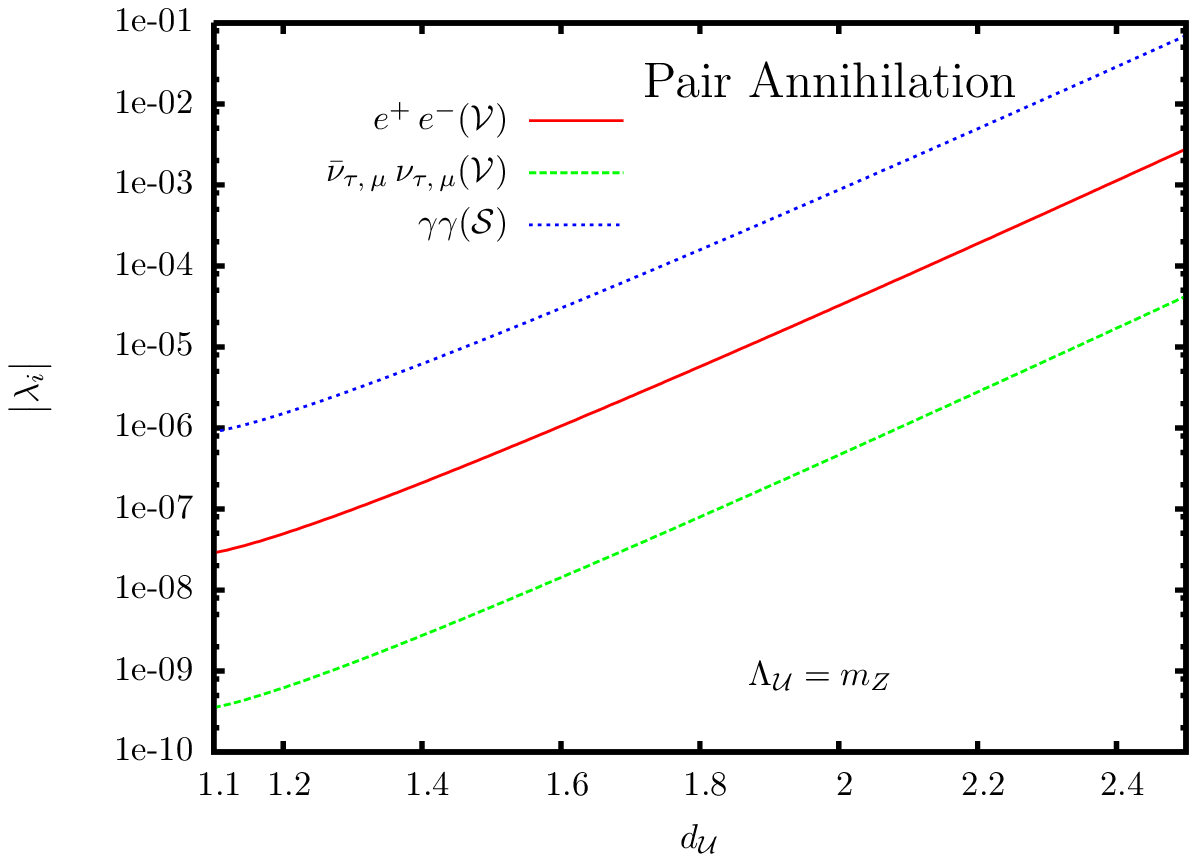}
\vskip -16 cm 
\caption{Contour on $d_{\cal U}$ and coupling $\bigl\vert \lambda_i\bigr\vert$ plane for $\Lambda_{\cal U}= m_Z$ for the unparticle emitting pair annihilation  processes}
\label{scatfig}
\end{center}
\end{figure}       
The  possible pair annihilation processes in the supernova core responsible for energy loss through the emission of unparticles are $\gamma\, \gamma \rightarrow {\cal U}$, $e^+\,e^- \rightarrow {\cal U}$ and  $\nu\,\bar \nu \rightarrow {\cal U}$. In the supernova core the electrons and electron neutrinos are degenerate with chemical potential typically of the order of 150-200 MeV whereas  muon and tau neutrinos are essentially non-degenerate. Further for $d_{\cal U}$ = 1 as mentioned in the section \ref{introduction}, the rate vanishes because of energy momentum conservation. Using the photons coupling to the scalar unparticles  given in equation (\ref{scalcoup}). 
The pair annihilation  cross- section is given by 
\begin{eqnarray}
\sigma_{av}\bigl(\gamma\,\gamma\rightarrow {\cal U}\bigr)&=& \frac{1}{8}\,\, \left(\frac{\bigl\vert\lambda_0^\gamma\bigr\vert}{\Lambda_{\cal U}^{d_{\cal U}}}\right)^2\,\, {\cal A }_{d_{\cal U}}\,\, s^{d_{\cal U}-1}
\end{eqnarray}
and the energy  loss is calculated to be
\begin{eqnarray}
\stackrel{\bf .}{\epsilon}_{\cal U}^\gamma &=& \frac{2^{(2\,d_{\cal U}+1)}}{d_{\cal U}+1 }\,{\cal A}_{d_{\cal U}}\,\left(\frac{\bigl\vert\lambda_0^\gamma\bigr\vert}{\Lambda_{\cal U}^{d_{\cal U}}}\right)^2 \,\, \frac{T^{(2\,d_{\cal U}+5)}}{(2\,\pi)^4}\zeta (d_{\cal U}+3)\,\Gamma (d_{\cal U}+3) \Gamma (d_{\cal U}+2)
\zeta (d_{\cal U}+2)
\end{eqnarray}
For $e^+\, e^-\rightarrow {\cal U} $, we have the contribution from the vector unparticle operator 
\begin{eqnarray}
\sigma_{av}^{e^+e^-}\bigl(e^+\, e^-\rightarrow {\cal U}\bigr)&=& \frac{1}{2}\,\,\left(\frac{ \bigl\vert\lambda_1^{\cal V}\bigr\vert}{\Lambda_{\cal U}}\right)^2\,\, {\cal A }_{d_{\cal U}}\,\,\left( \frac{s}{\Lambda_{\cal U}^2}\right)^{d_{\cal U}-2}
\end{eqnarray}
and for $\nu\,\bar\nu \rightarrow {\cal U}$ we have 
\begin{eqnarray}
\sigma_{av}\bigl(\nu\, \bar\nu\rightarrow {\cal U}\bigr)&=& 2\,\,\sigma_{av}\bigl(e^+\, e^-\rightarrow {\cal U}\bigr)
\end{eqnarray}
The energy loss rate is given by 
\begin{eqnarray}
\stackrel{\bf .}{\epsilon}_{\cal U}^{e^+e^-} &=& \frac{2^{(2\,d_{\cal U}-2)}}{(8\,\pi )^4\,\, d_{\cal U} }\,{\cal A}_{d_{\cal U}}\,\left(\frac{\bigl\vert\lambda_1^{\cal V}\bigr\vert }{\Lambda_{\cal U}^{d_{\cal U}-1}}\right)^2 \,\, T^{(2\,d_{\cal U}+3)}\,\left[\prod_{i=1}^2\int\frac{x_i^{d_{\cal U}}\, dx_i}{e^{x_i+(-\, 1)^i\, y}+1}\right]\,\, \bigl(x_1+x_2\bigr)
\end{eqnarray}  
where $y=\mu_F/ T$. For $\nu_\mu$ and $\nu_\tau $ we can take $y$ to be zero and hence the energy loss rate becomes 
\begin{eqnarray}
\stackrel{\bf .}{\epsilon}_{\cal U}^{\nu_\mu ,\,\nu_\tau} &= & \frac{2^{(2\, d_{\cal U}+1)}}{(2\,\pi )^4\,\, d_{\cal U} }\,{\cal A}_{d_{\cal U}}\,\left(\frac{\bigl\vert\lambda_1^{\cal V}\bigr\vert }{\Lambda_{\cal U}^{d_{\cal U}-1}}\right)^2 \,\, T^{(2\,d_{\cal U}+3)}\,
\zeta (d_{\cal U}+2)\,\Gamma (d_{\cal U}+2)\,\zeta (d_{\cal U}+1) \Gamma (d_{\cal U}+1)\nonumber\\
&& \,\,\,\, \times\,\,\, \left[1-\frac{1}{2^{(d_{\cal U}+1)}}\right]^{-1}\, \left[1-\frac{1}{2^{d_{\cal U}}}\right]^{-1}\nonumber\\
&=&\left(\frac{\bigl\vert\lambda_1^{\cal V}\bigr\vert}{\bigl\vert\lambda_0^\gamma\bigr\vert }\right)^2\, \frac{1}{4}\, \frac{1}{(d_{\cal U}+2)\,\,d_{\cal U} }\,\left(\frac{\Lambda_{\cal U}}{T}\right)^2\,\, \frac{\zeta (d_{\cal U}+1)}{\zeta (d_{\cal U}+3)}\,\, \left[1-\frac{1}{2^{(d_{\cal U}+1)}}\right] \left[1-\frac{1}{2^{d_{\cal U}}}\right]\,\,\stackrel{\bf .}{\epsilon}_{\cal U}^\gamma
\end{eqnarray}
For the processes induced by the  tensor unparticle operators, the corresponding energy loss rate for pair annihilation has been calculated in the reference \cite{lewis}.
\par As discussed above, the limits on $d_{\cal U}$ = 1 case can be obtained by considering the pair annihilation process through $e^+\,e^-\rightarrow {\cal U}\,\gamma $. We first consider the  emission of vector unparticle
\begin{equation}
\frac{\lambda_1^{\cal V}}{\Lambda_{\cal U}^{d_{\cal U}-1}}\,\, \bar f \,\gamma_\mu\, f\, {\cal O}^\mu_{\cal U}\, .
\end{equation}
There are two Feynman diagrams in the $u$ and $t$ channel. The matrix element squared is given by
\begin{eqnarray}
\bigl\vert{\cal M}\bigr\vert^2&=& 2\, \bigl( 4\,\pi\alpha_{em}\bigr)\, \left(\frac{\bigl\vert\lambda_1^{\cal V}\bigr\vert}{\Lambda_{\cal U}^{d_{\cal U}-1}}\right)^2 \,\,\, \frac{2\, s\, P^2+u^2+t^2}{u\,\,t}
\end{eqnarray}
and the cross section is given as 
\begin{eqnarray}
\sigma_{av}&=& \frac{1}{2\,s}\, \frac{{\cal A}_{d_{\cal U}}}{16\,\pi^3}\,\int\bigl\vert{\cal M}\bigr\vert^2\, \bigl(P^2\bigr)^{d_{\cal U}-2}\,\, E_\gamma\, dE_\gamma \, d\Omega_\gamma \,\, \Theta\bigl(P^2\bigr)\, \Theta\bigl(P_0\bigr) 
\end{eqnarray}
In the limit  $d_{\cal U}$ = 1
\begin{eqnarray}
{\rm Lt.}_{\,\,\,\, d_{\cal U} \rightarrow 1+} \,\,\,\, {\cal A}_{d_{\cal U}}\,\,\bigl(P^2\bigr)^{d_{\cal U}-2}\,\,\, \Theta\bigl(P^2\bigr)=2\,\pi\,\delta (P^2)= \frac{\pi}{\sqrt{s}}\,\,\delta \left(E_\gamma-\frac{\sqrt{s}}{2}\right)
\end{eqnarray}
Therefore the cross section becomes
\begin{eqnarray}
\sigma_{av}^{\cal V}&=&\alpha_{em}\, \bigl\vert{\lambda_1^{\cal V}}\bigr\vert^2\,\,\frac{1}{8\,\pi}\,\int \frac{1}{s} \,\,\left(\frac{u}{t}+\frac{t}{u}\right)\,\, d\Omega_\gamma\nonumber\\&=& \alpha_{em}\, \bigl\vert{\lambda_1^{\cal V}}\bigr\vert^2\,\,\frac{1}{4}\,\,\frac{1}{s} \,\,\left(1+\ln\frac{s}{m_e^2}\right)  \label{guxsc} 
\end{eqnarray}
The emissivity is given as
\begin{eqnarray}
\stackrel{\bf .}{\epsilon}^{\cal V} &=& \frac{2\,\alpha_{em}\,\bigl\vert{\lambda_1^{\cal V}}\bigr\vert^2}{4\,(2\,\pi )^4}\,\, T^5\,\, \left[\prod_{i=1}^2\int_0^\infty \frac{dx_i}{e^{x_i+(-\,1)^i\,y}+1}\right]\,\,\int_0^{\beta }\,\, \left(x_1+x_2-\frac{\sqrt{Z}}{2}\right) \,\left( 1+ \ln \frac{Z}{\alpha }\right)\,\, dZ\nonumber\\
&&{\rm where } \,\,\, Z=\frac{s}{T^2};\,\, \beta=4\,x_1\,x_2\,; \,\,\, {\rm and}\,\,\, \alpha = \left(\frac{m_e} {T}\right)^2\nonumber\\
&=& \frac{2\alpha_{em} \bigl\vert{\lambda_1^{\cal V}}\bigr\vert^2}{4(2\pi )^4}T^5 \left[\prod_{i=1}^2\int_0^\infty \frac{dx_i}{e^{x_i+(-\,1)^iy}+1}\right] \left[\ln\alpha^{-\beta\left(x_1+x_2\right)} + \frac{\beta^{3/2}}{9} ( \ln\alpha^3-1 )+\left\{x_1+x_2-\frac{\sqrt{\beta}}{3}\right\}\ln\beta^\beta\right]\nonumber\\
\label{gu}
\end{eqnarray}
\par The leading term in the cross - section for scalar, pseudo-scalar and axial vector couplings has same behaviour  namely 
\begin{eqnarray}
\sigma_{\rm av} \approx \frac{1}{s}\,\, \ln\frac{s}{m_e^2}
\end{eqnarray}
and the magnitude is roughly half of the vector case. The energy loss is, therefore given by equation (\ref{gu}) within a factor of 2.
For the tensor unparticle operator given in equation (\ref{tenscoup}), the cross - section for the process  $e^+\,e^-\rightarrow {\cal U}\,\gamma $ is given by
\begin{eqnarray}
\sigma_{av}^{\cal T}&=&\frac{1}{9}\,\,\, \alpha_{em}\,\,\, \left(\frac{\bigl\vert\lambda^{\cal T}\bigr\vert}{\Lambda_{\cal U}}\right)^2
\end{eqnarray}
and the emissivity calculated in the limit $d_{\cal U} \rightarrow 1$ is given as
\begin{eqnarray}
\stackrel{\bf .}{\epsilon}^{\cal T}(d_{\cal U}=1) &=& \frac{2\,\, \alpha_{em}}{9\,\,\, (2\,\pi )^4}\,\,\left(\frac{\bigl\vert\lambda^{\cal T}\bigr\vert}{\Lambda_{\cal U}}\right)^2\, T^7\, \left[\prod_{i=1}^2\int_0^\infty \frac{dx_i}{e^{x_i+(-\,1)^i\,y}+1}\right]\,\left[8\, (x_1+x_2) \, x_1^2\, x_2^2- \frac{\bigl(4\, x_1\, x_2\bigr)^{5/2}}{5}\right]\nonumber\\
\end{eqnarray}
\section{Numerical Estimation and Discussion}
\label{discuss}
Observation of neutrino flux from IMB and  Kamiokande established that the most of the energy released during supernovae explosion was carried away by neutrinos. This observation has been used to place constraints on  new sources of energy loss by demanding the energy loss per unit volume per second does not exceed $\stackrel{\bf .}{\epsilon}_{SN} \approx 3 \times 10^{33}$ ergs cm$^{-3}$ s$^{-1} = 9.45\times 10^{-15}\,\, {\rm MeV}^5$. Using this upper bound on the energy loss rate, we  put bounds on the parameters involved in the unparticle  theories. 
\par For calculating the energy loss rate due to nucleon bremsstrahlung and pair annihilation processes from the supernova core, we take the core temperature $T $ to be 30 MeV, effective nucleon mass
$m_n^\star\approx 0.8 \,\, m_n $ where $m_n$ is mass of the nucleon and neutron Fermi momentum $p_F\approx $ 515 $\rho_{15}^{1/3}$ MeV. We have used $p_F$ = 345 MeV for $\rho\approx 3\times 10^{14}$ gms per cm$^3$.  In Figure \ref{bremfig} we have provided the contours on the $d_{\cal U}$ and $\bigl\vert \lambda\bigr\vert$ plane by restricting the energy loss rate due to unparticle emissivity to be 3 $\times $ 10$^{33}$   ergs cm$^{-3}$ s$^{-1}$ induced by scalar, pseudo-scalar, vector and axial vector unparticle operators. In these calculations the energy scale $\Lambda_{\cal U}$ has been normalized to  $m_Z$. 
\par The bounds on pseudo-scalar and pseudo-vector interactions as can be seen from figure \ref{bremfig} are most restrictive and we obtain $\bigl\vert\lambda_{0,1}^{\cal P}\bigr\vert\leq 4\times 10^{-11}$ for $d_{\cal U}=1$. The corresponding bounds on scalar and vector couplings are $\bigl\vert\lambda_{0,\, 1}^{\cal S,\,V}\bigr\vert\leq 7\times 10^{-11}$.
\par For the purpose of calculating energy loss rate from pair annihilation processes, the electron chemical potential in the supernova core is taken to be $\mu_e\approx $ 345 MeV. Based on the similar procedure as mentioned above we give the  corresponding contours from pair annihilation processes in figure \ref{scatfig}. 
We find that the contribution to  the emissivity  from the $\gamma\gamma$ annihilation process from the scalar unparticles is identical to that of the tensor unparticle as given in the reference \cite{lewis}.
\par As discussed in the text, we utilize the pair annihilation processes  $e^+\,e^-\rightarrow {\cal U}\,\gamma $  for constraining the dimensionless coupling for $d_{\cal U}$ = 1. Following equation (\ref{gu}) we find
\begin{eqnarray}
\stackrel{\bf .}{\epsilon}&\approx&15.4\, \bigl\vert{\lambda_1^{\cal V}}\bigr\vert^2\,\,\, {\rm MeV}^5
\end{eqnarray}
Therefore SN 1987A constraints the vector dimensionless coupling  $\bigl\vert\lambda_1^{\cal V}\bigr\vert \leq  2.5\times 10^{-8}$. The dimensionless coupling  $\bigl\vert\lambda^{\cal T}\bigr\vert$ is constrained from SN 1987A as
\begin{eqnarray}
\bigl\vert\lambda^{\cal T}\bigr\vert &\leq & 7.26 \times 10^{-4} \,\,\left[\frac{ \Lambda_{\cal U}}{1\,\, {\rm TeV}}\right]^{2}\, .
\end{eqnarray}
 For $\Lambda_{\cal U}=m_Z$ we find for $d_{\cal U}=1$,  $\bigl\vert\lambda^{\cal T}\bigr\vert \leq  6\times 10^{-6}$.  Another interesting feature worth mentioning is that the cross section and hence the energy loss rate for the vector unparticle interaction as shown in equations (\ref{guxsc}) and (\ref{gu}) are independent of the energy scale $\Lambda_{\cal U}$. 
\section*{Acknowledgement} 
The authors thank Debajyoti Choudhury for helpul discussions.

\end{document}